\documentclass{iau}
\usepackage{graphicx}

\def\DJ{\mbox{\raise0.3ex\hbox{-}\kern-0.4em D}}

\title[Secular evolution of asteroid families] %% give here short title %%
{Secular evolution of asteroid families: \\ the role of Ceres}

\author[B. Novakovi\'c, G. Tsirvoulis, S. Mar\`o, V. \DJ o\v sovi\' c \& C. Maurel]   %% give here short author list %% 
{Bojan Novakovi\'c$^1$
Georgios Tsirvoulis$^2$
Stefano Mar\`o$^3$ \\
Vladimir \DJ o\v sovi\' c$^1$
\and Clara Maurel$^4$}

\affiliation{$^1$Department of Astronomy, Faculty of Mathematics, University of
Belgrade, Studentski trg 16, 11000 Belgrade, Serbia \\ email: {\tt bojan@matf.bg.ac.rs} \\[\affilskip]
$^2$Astronomical Observatory, Volgina 7, 11060 Belgrade 38, Serbia  \\[\affilskip]
$^3$Dipartimento di Matematica, Universit\`a di Pisa, Largo B. Pontecorvo 5, 56127 Pisa, Italy \\[\affilskip]
$^4$Institut Sup\'{e}rieur de l'a\'{e}ronautique et de l'Espace
(ISAE-Supa\'{e}ro), University of Toulouse, 31055 Toulouse Cedex 4, France}

\pubyear{2016}
\volume{318}  
\pagerange{000--000}
\jname{Asteroids: New Observations, New Models}
\editors{S. Chesley, A. Morbidelli, R. Jedicke \& D. Farnocchia, eds.}
\begin{document}

\maketitle

\begin{abstract}

We consider the role of the dwarf planet Ceres on the secular dynamics of the asteroid main belt. 
Specifically, we examine the post impact evolution of asteroid families due to the interaction of 
their members with the linear nodal secular resonance with Ceres. First, we find the location of 
this resonance and identify which asteroid families are crossed by its path. Next, we summarize our 
results for three asteroid families, namely (1726) Hoffmeister, (1128) Astrid and (1521) Seinajoki 
which have irregular distributions of their members in the proper elements space, indicative of the 
effect of the resonance. We confirm this by performing a set of numerical simulations, showcasing 
that the perturbing action of Ceres through its linear nodal secular resonance is essential to 
reproduce the actual shape of the families.


\end{abstract}

\firstsection % if your document starts with a section,
\section{Introduction}

An asteroid family is a group of asteroids orbiting the Sun in very similar orbits, 
after being broken apart, from a single parent body, by a collision with another small 
solar system object. These collisionally formed groups have attracted a lot of attention of 
solar system scientists because they can provide information on many asteroid-related processes 
(\cite[Cellino et al. 2009]{cellino2009}). Examples include information about the internal
structure of their parent bodies (\cite[Cellino et al. 2002]{cellino2002}), 
time-scale of space weathering process (\cite[Vernazza et al. 2009]{vernazza2009}), 
the formation of binary asteroids (\cite[Michel et al. 2001, 
Durda et al. 2004]{michel2001,durda2004}), the collisional history of the main asteroid belt 
(\cite[Cibulkov{\'a} et al. 2014]{cibulkova2014}), as
well as many other asteroid-related subjects.

As the relative ejection velocities of the fragments are small compared to the parent 
body's orbital speed, it is expected that the orbital elements of each member at breakup 
time are very close to those of the parent. However, families evolve significantly since 
the epoch of their formation as a consequence of different processes and perturbatations, such as:
chaotic diffusion (\cite[Nesvorny et al. 2002, Novakovi\'c 2010, Novakovi\'c et al. 2010a]{nes2002,nov2010,nov2010a}), 
semi-major axis drift due to the Yarkovsky effect (\cite[Farinella \& Vokrouhlicky 1999, Bottke et al. 2001, Spoto et al. 2015]{farvok1999,bottke2011,spoto2015}), secondary collisions (\cite[Marzari et al. 1999, 
Milani et al. 2014]{marzari1999,milani2014}), non-destructive collisions (\cite[Dell'Oro \& Cellino 2007]{delloro2007}), and close encounters with massive asteroids (\cite[Carruba et al. 2003, Novakovi\'c et al. 2010b]{carruba2003,nov2010b}).

\cite[Novakovi{\'c} et al. (2015)]{Novakovic2015} have recently shown that the linear nodal secular 
resonance with (1) Ceres $\nu_{1c}=s-s_{c}$ (here $s$ and $s_{c}$ denote the proper frequencies of
ascending node of an asteroid and Ceres respectively), is mainly responsible for the evolution
of the Hoffmeister asteroid family. That was the first time  compelling evidence for the  
orbital evolution of small bodies caused by a secular resonance with an asteroid has been found. 

The fact that asteroid families evolve over time, poses a problem of reconstructing the 
original collisional outcome from the knowledge of the current properties of 
family members. Thus, it is of extreme importance to identify and understand all the 
mechanisms that may result in family evolution.
The aim of this work is to study the importance of the $\nu_{1c}$ resonance on the long-term 
dynamics of asteroid families.

\section{The $\nu_{1c}$ resonance}

In order to study the importance of the $\nu_{1c}=s-s_c$ secular resonance on the 
orbits of asteroids, it is essential to first locate which parts of the main asteroid belt 
are affected, i.e. to determine the location of this resonance in the proper orbital
elements space.

The location of a given secular resonance could be found analytically
(\cite[Kne\v zevi\' c et al. 1991]{Knezevic1991}), but this approach 
has some limitations, especially for high
eccentricity and inclination regions of the main-belt. For this reason we
use here a different, numerical approach, based on the proper elements of main-belt
asteroids (\cite[Kne{\v z}evi{\'c} and Milani 2003]{Knezevic2003}). 
The asteroids that are currently in resonance are
those for which the critical angle $\sigma=\Omega-\Omega_c$ librates, and
consequently they satisfy the resonant relation $s\simeq s_c$. Taking advantage of
this, we can find the location of the secular resonance by plotting only those
asteroids that satisfy this relation. 

A preceding step is to define which is the relevant width of the resonance, i.e. what is the maximum 
difference between the nodal frequencies of Ceres and the one of an asteroid, in order to consider an 
object to be likely resonant. \cite[Milani \& Kne{\v z}evi{\'c} (1994)]{Milani1994} used a tolerance of $2\,arcsec/yr$ for the strong 
$g-g_6$ secular resonance, and $0.5\, arcsec/yr$ for the weaker, fourth degree secular resonances such as $g+s-g_6-s_6$ for example. Other authors also use similar values of
$0.1-0.5\,arcsec/yr$ for different secular resonances (see e.g. \cite[Carruba et al. 2009; Mili{\'c} {\v Z}itnik and Novakovi{\'c} 2015]{carruba2009,ivana2015}). Based on this, and having in mind that
we should expect the $\nu_{1c}$ secular resonance to be comparatively weak, we chose a tolerance of
$0.2\,arcsec/yr$. This value is a reasonable balance between being large enough to identify the asteroid families that are crossed by the resonance, and at the same time keeping the plot clear. 

\begin{figure}[ht!]
% \vspace*{-2.0 cm}
\begin{center}
\includegraphics[width=0.85\textwidth]{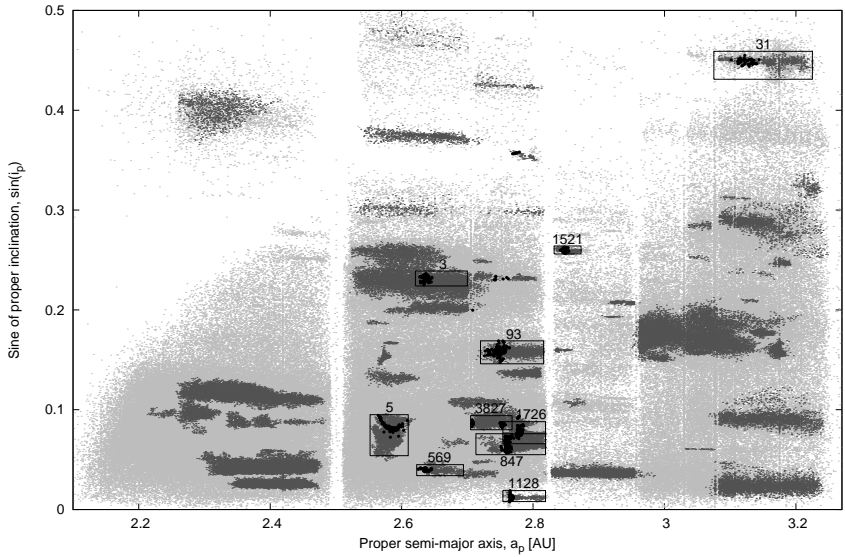} 
% \vspace*{-1.0 cm}
 \caption{Asteroid families affected by the $\nu_{1c}=s-s_c$ secular resonance, in the proper
semi-major axis versus sine of proper inclination plane $(a_p,\sin(i_p))$. In light gray dots we
show all main-belt asteroids while in dark gray dots only asteroids belonging to
families. The black points show asteroid family members that are also in resonance
$(\mid s-s_c \mid <0.2)$. The black boxes highlight the asteroid families that have
resonant members.}
   \label{fig:res_fam}
\end{center}
\end{figure}

The result is shown in Figure \ref{fig:res_fam}, where we plotted the proper
semi-major axis versus the sine of proper inclination $(a_p,\sin(i_p))$ of the main-belt
asteroids. For clarity, we plotted in black only those resonant
asteroids that belong to asteroid families (highlighted in dark gray), which helps
us assess which families are crossed by the resonance. From this we were able to
identify 10 asteroid families, using the classification of \cite[Milani et al. (2014)]{milani2014},
that have a considerable number of likely resonant members. These families are: (3) Juno,
(5) Astraea, (31) Euphrosyne, (93) Minerva, (569) Misa, (847) Agnia, (1128) Astrid,
(1521) Seinajoki (also known as (293) Brasilia), (1726) Hoffmeister and (3827) Zdenekhovsky. 

In this paper we wish to show the effect of the $\nu_{1c}$ secular resonance on the
asteroid families where it is more prominent. We remind here that this secular
resonance, which involves only the precession frequency $s$ of the ascending node
$(\Omega)$, is causing perturbations mainly in the orbital inclination.
Therefore we will summarise our results for the cases of (1726)~Hoffmeister, 
(1128)~Astrid and (1521)~Seinajoki because these
three families are the ones that suggest, based on their shape on the
$(a_p,\sin i_p)$ plane, that the secular resonance with Ceres has caused significant
evolution of their members' orbits. 

\section{Affected asteroid families}

In this section we present our results of dynamical evolution of the Hoffmeister, 
Astrid and Seinajoki families, focusing on the role of Ceres.
In the following we first present our methodology, and then the results for each family,
on a case-by-case basis.

\subsection{Methodology}

The main idea behind the procedure used here is to
reproduce the orbital evolution of the family members since the breakup event.
Practically, we are simulating the evolution in time of the orbits of family members
by using numerical integrations. These integrations are performed
using two different dynamical models. Both models include the gravitational effects 
of the Sun and the four outer planets (from Jupiter
to Neptune) and also account for the Yarkovsky thermal force. The second model differs
from the first one, in that it also takes into account the most massive asteroid (1) Ceres 
as a perturbing body. Therefore, a comparison of the outcomes obtained under these 
two models should allow us to characterize the exact role played by Ceres for 
the dynamical evolution of these three asteroid families. 

For the integrations we employed the \emph{ORBIT9} integrator embedded in the multipurpose \emph{OrbFit} package (available from http://adams.dm.unipi.it/orbfit/). The settings for the Yarkovsky effect are
made by assigning to each particle a random value from the interval $\pm (da/dt)_{max}$, with $(da/dt)_{max}$ being the estimated maximum of the semi-major axis drift speed caused by the Yarkovsky force. 
The maximum drift speed is derived from the model of the Yarkovsky effect developed by \cite[Vokrouhlick{\'y} (1998,1999)]{vok1998,vok1999}, and assuming appropriate thermal parameters. 

In particular, for two dark, $C$-type families, namely the Hoffmeister and Astrid, 
we adopt values of $\rho_{s}$ = $\rho_{b}$ = 1300~$kg~m^{-3}$
for the surface and bulk densities (\cite[Carry 2012]{carry2012}), $\Gamma$ = 250~$J~m^{-2}~s^{-1/2}~K^{-1}$
for the surface thermal inertia (\cite[Delb\' o \& Tanga 2009]{delbo2009}), and $\epsilon$ = 0.95 
for the thermal emissivity parameter. 
For each family, we used its mean geometric albedo calculated using data from 
\cite[Masiero et al. (2011)]{wise}.
In this way we found that for a body of $D=1$~km in diameter $(da/dt)_{max}$
is about $4.5 \times 10^{-4}$~AU/Myr. 

For the Seinajoki, which seems to be an $S$-type family, we adopt values of $\rho_{b}$ = 2300~$kg~m^{-3}$
for the bulk density, and $\Gamma$ = 125~$J~m^{-2}~s^{-1/2}~K^{-1}$
for the surface thermal inertia. These parameters, at location of the Seinajoki family, give
$(da/dt)_{max}$ of $3.0 \times 10^{-4}$~AU/Myr, for a body of $D=1$~km in diameter.

Finally, as the Yarkovsky effect scales as $ \propto 1/D$, to each test particle we attribute
a diameter selected in such a way that a size-frequency distribution (SFD) of the test particles
resembles SFD of the real family members. In the case of the Hoffmeister family the number of 
test particles used in simulations was equal to the number of asteroids identified as the family 
members. For the Astrid and Seinajoki families, the number of real family members were too small
to reliably simulate evolution of these families. Thus, in these two cases we extrapolate 
the SFDs of real families down to smaller sizes till the desired number of 2000 test particles. 
 
For each family, the orbits of the test particles are propagated for 150~Myr, starting
from the expected distribution of family members immediately after the breakup event.
The integration time span adopted here should be long enough to reveal the possible effect of Ceres. 
Next, following \cite[Kne{\v z}evi{\'c} and Milani (2000)]{knemil2000}, the time series of mean orbital elements, obtained removing the short-periodic perturbations
from the instantaneous osculating elements, are produced using on-line
digital filtering. Finally, for each particle we compute the 
proper elements for consecutive intervals of 10~Myr. These steps make it possible
to study the evolution of the families directly in the space of proper elements.

\subsection{The Hoffmeister family}

The (1726) Hoffmeister asteroid family is located in the middle of the main asteroid belt,
between $2.75$ and $2.82$~AU (see Figure~\ref{fig:res_fam}). The orbits of its members are 
characterized by low proper orbital eccentricities, as well as low proper orbital inclinations. 
\cite[Novakovi{\'c} et al.(2015)]{nov2015} has recently demonstrated that the unusual shape
of the Hoffmeister family is a result of the perturbation related to the nodal secular resonance 
with Ceres $\nu_{1c}=s-s_c$. Here we follow essentially the same steps to show this effect.

\begin{figure}[ht!]
% \vspace*{-2.0 cm}
\begin{center}
\includegraphics[width=0.85\textwidth]{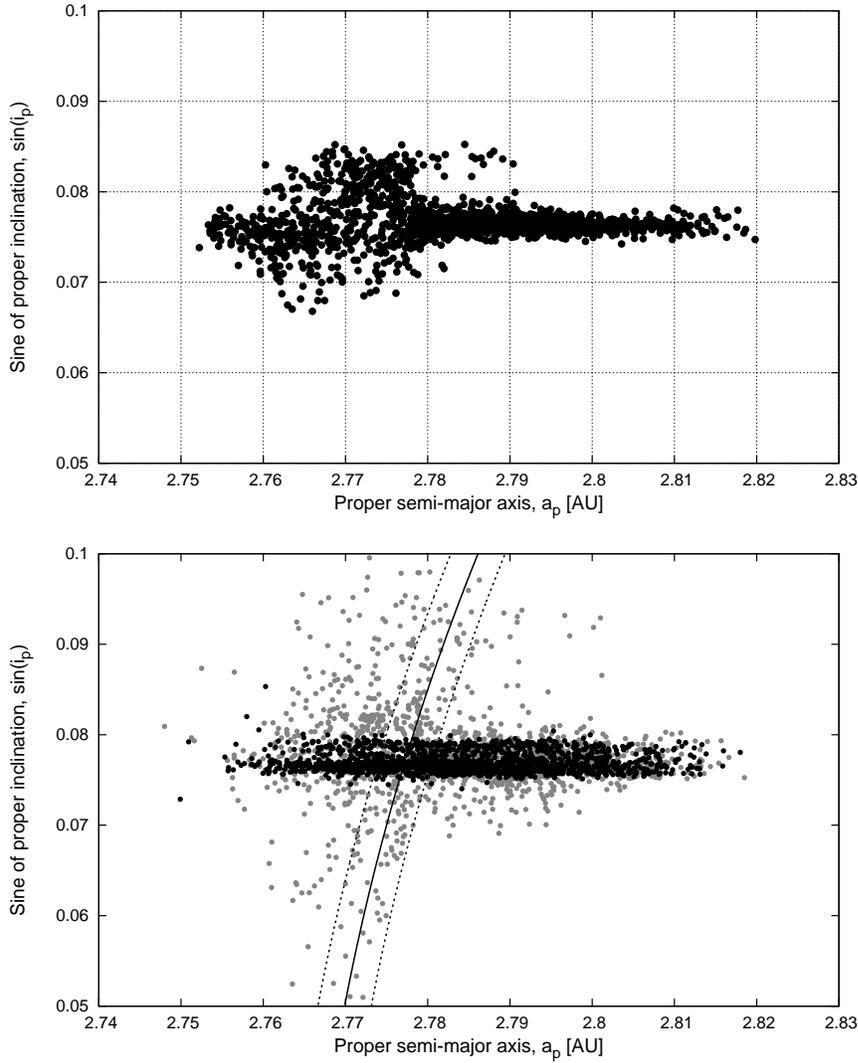} 
% \vspace*{-1.0 cm}
 \caption{\textit{Top}: The distribution of the nominal Hoffmeister family members in the 
 proper semi-major axis versus the proper inclination plane; \textit{Bottom}: The distribution
 of the test particles after $150$~Myr of the evolution. The black points represent
 particles integrated within the model without Ceres, while the gray points denote
 particles simulated using the dynamical model that also includes Ceres as a perturbing body.
 The solid and two dashed curves mark the center and the borders of the $\nu_{1c}=s-s_c$ resonance, respectively.}
\label{fig:1726}
\end{center}
\end{figure}

The effect of Ceres can be very easily appreciated from the integrations of test particles
performed within the two different dynamical models. The particles integrated within the
model without Ceres spread only in the orbital semi-major axis due to the Yarkovsky effect,
with practically no evolution at all in the orbital inclination (see bottom panel in 
Figure~\ref{fig:1726}). Thus, the distribution of test particles at the end of the integration
time span obviously cannot explain the shape of the Hofmeister family as seen in the ($a_p,\sin i_p$) plane
(see top panel in Figure~\ref{fig:1726}).

The behaviour of test particles changes significantly when Ceres is added to the model.
The striking difference observed in these runs is the large dispersion of orbital inclinations.
In the latter case, the distribution of test particles after 150~Myr of evolution is very similar
to the distribution of real family members (Figure~\ref{fig:1726}). These results undoubtedly
confirm that perturbations induced by Ceres are responsible for the changes in the inclination.

\subsection{The Astrid family}

The (1128) Astrid asteroid family is located in the middle belt, more precisely, its position in 
the proper elements space is characterized by proper semi-major 
axes varying from $2.75$ to $2.82$~AU, proper eccentricity lower than $0.055$, and 
proper inclination of approximately $0.7$~degrees. The family is well separated from any
other group in the main belt, as well as from the local background population.
An analysis of the zone surrounding the family in the $(a_p, \sin i_p)$ plane
reveals that the family is isolated as only a few background objects
are present. 

\begin{figure}[ht!]
% \vspace*{-2.0 cm}
\begin{center}
\includegraphics[width=0.85\textwidth]{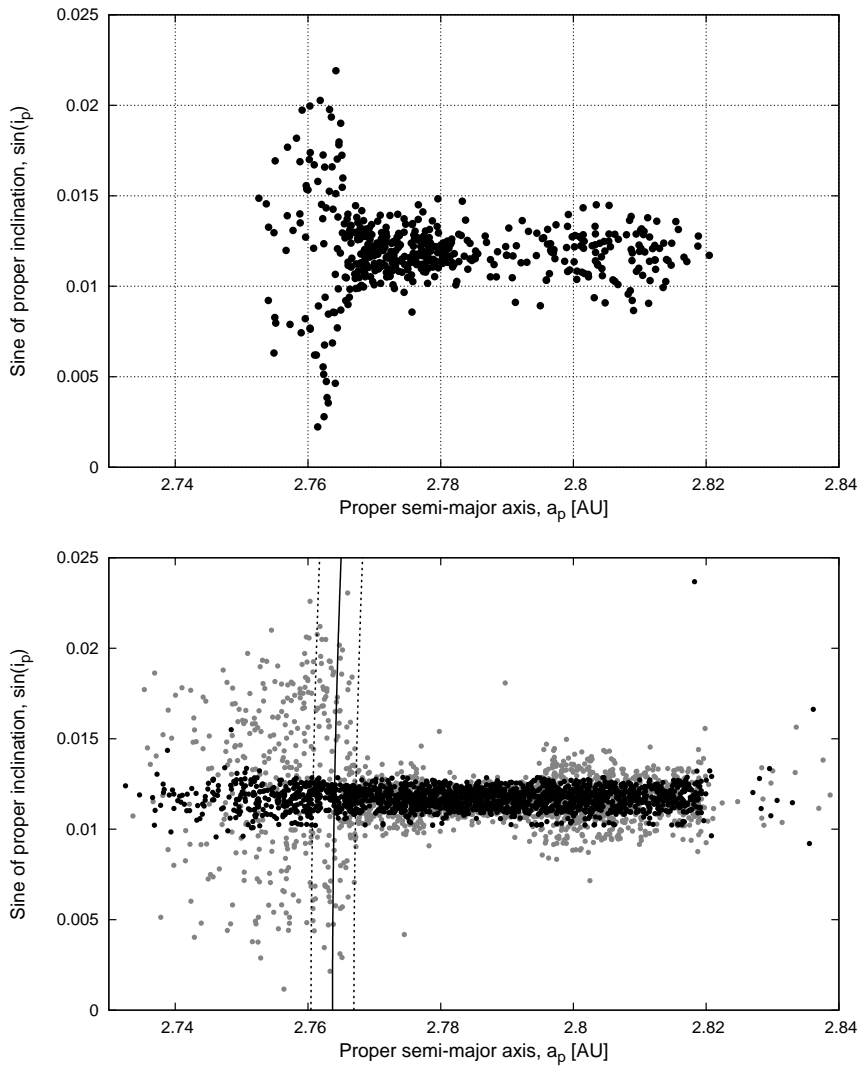} 
% \vspace*{-1.0 cm}
 \caption{The same as in Figure~\ref{fig:1726}, but for the (1128) Astrid asteroid family.}
\label{fig:1128}
\end{center}
\end{figure}

The distribution of Astrid family members projected on the $(a_p,\sin i_p)$ plane (Figure~\ref{fig:1128}) shows a pattern similar to the case of the Hoffmeister family. 
For smaller values of $a_p$, the spread in orbital inclinations is notably bigger than 
at larger values of $a_p$. Moreover, a similar \textit{lobe} is also present in the right
side of the family, even if it is not so prominent.

Knowing that the Astrid family is also crossed by the $\nu_{1c}$ resonance with Ceres,
we hypothesized that this resonance is responsible for the dispersion of the family
in inclination. Thus, to validate our conjecture we numerically simulated the evolution 
of the family under the two different dynamical models.

As in the case of the Hoffmeister family, if we include in our model only the giant
planets and the Yarkovsky force, we only see the dispersion along the semi-major axis 
due to the Yarkovsky effect (see bottom panel in Figure~\ref{fig:1128}). Hence, it is 
clear that we cannot reproduce the shape of the family, and that the dynamical evolution 
subject only to the perturbations by the Yarkovsky force and the giant planets is not 
enough to explain the evolution of this family.

On the other hand, when Ceres is included in the dynamical model,
the difference with respect to the previous case is evident, and the
shape of the real family is well reproduced. This can be appreciated from
the bottom panel shown in Figure~\ref{fig:1128}, where gray points represent 
the state of the test particles after $150$~Myr of evolution within the model with Ceres.
Note that the spread in inclination occurs when a
particle enters into the relevant resonance. Certainly, the Yarkovsky effect is
still efficient on particles in the $\nu_{1c}$ resonance with Ceres, so that
they can escape from this resonance but with a significantly different
value of proper inclination.  

We conclude noting that including Ceres
we can also reproduce the small lobes in inclination on the right side
of the family. These are not reproduced by the first model so that we
can conjecture that another resonance involving Ceres is affecting the
family in that zone. However, the identification of this week resonance
can be a tricky thing, because usually 
there may be more than one possible solution. Our preliminary investigation
suggests that this small effect in inclination of the Astrid family members
may be consequence of the $s-s_{c}+g_{c}-2g_{6}+g_{5}$ secular resonance. 
Still, further analysis is needed to completely clarify this issue.

\subsection{The Seinajoki family}

The (1521) Seinajoki asteroid family is situated in the outer 
part of the main asteroid belt, at proper semi-major axis of from about $2.83$ to $2.9$~AU. 
Its family members have an average proper orbital inclination of about $15$ degrees, 
and proper orbital eccentricity around $0.12$.

\begin{figure}[ht!]
% \vspace*{-2.0 cm}
\begin{center}
\includegraphics[width=0.85\textwidth]{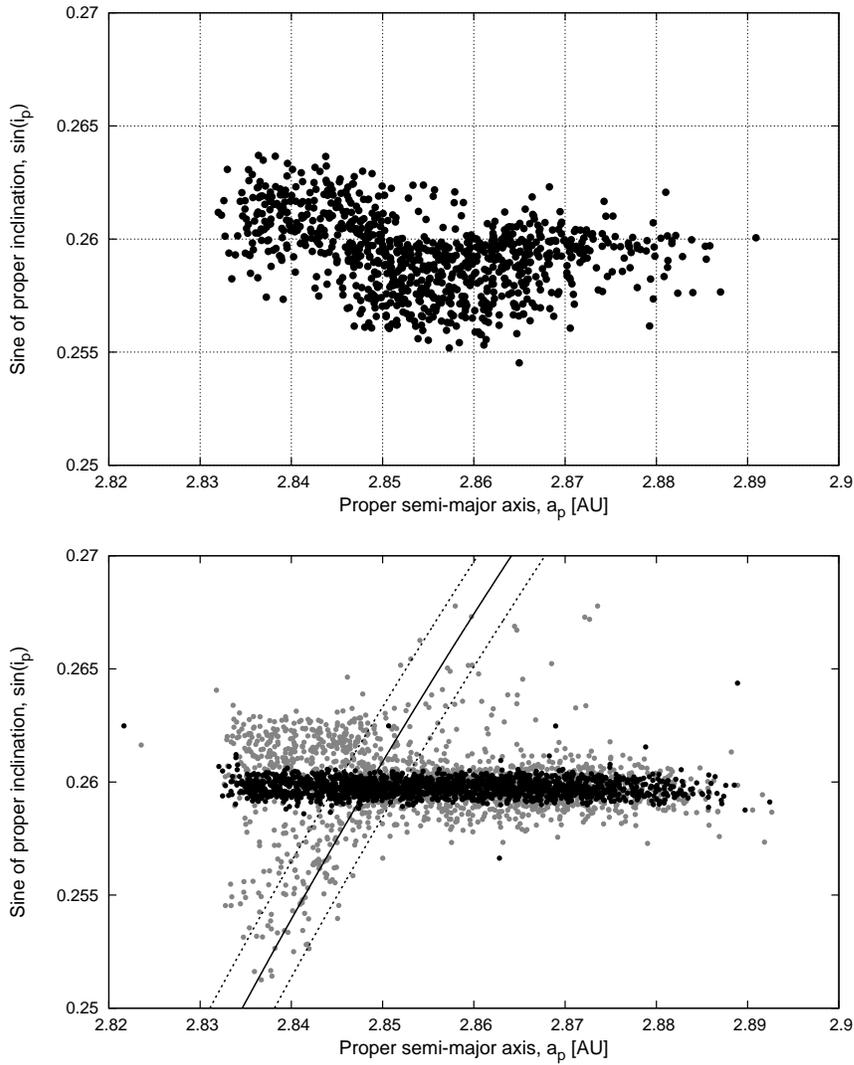} 
% \vspace*{-1.0 cm}
 \caption{The same as in Figure~\ref{fig:1726}, but for the (1521) Seinajoki asteroid family.}
\label{fig:1521}
\end{center}
\end{figure}

The shape of the Seinajoki family in the ($a_p, \sin i_p$) plane is also a bit strange,
as it is the case for the other two families analyzed here. Indeed, the dispersion of orbital 
inclinations at smaller semi-major axes exceeds that at larger $a_p$. 
However, a careful inspection of the distribution of family members in this plane  
(shown in top panel in Figure~\ref{fig:1521}) reveals an important difference. 
In the case of the Seinajoki family, the typical orbital inclinations at smaller
$a_p$ are systematically higher than those at larger $a_p$. Interestingly,
this offset corresponds very well to the location of the $\nu_{1c}$ secular resonance 
with Ceres.

The results obtained by simulating the evolution of test particles under two dynamical models
(with and without Ceres) show that Ceres is playing a very important role
in the dynamical evolution of this family. As in the two previous cases, without Ceres
included in the model, dispersion of proper orbital inclinations does not occur at all
(see bottom panel in Figure~\ref{fig:1521}). 

With Ceres included in the model evolution of test particles is notably different, and a
significant dispersion of inclinations is observed. Moreover, even the offset between
the orbital inclinations at smaller and larger $a_p$ is reproduced well. Once again this
suggests that Ceres is responsible for the shape of the family that we see today.
 
Finally, let us also mention that in the case of the Seinajoki family, even with Ceres 
included in the dynamical model, our simulations did not reproduce the current shape of the 
family that well, as in the case of the Hoffmeister and Astrid families. In this respect,
we would like to recall that the proper inclinations as well as the proper eccentricities
of the Seinajoki family members are significantly larger than for the other two families. Thus, 
although further investigation along these lines is beyond the scope of this work, we 
speculate that an explanation for this discrepancy may be an incomplete dynamical model
used here. Specifically, perturbations caused by the inner planets may be relevant for the
dynamics of the asteroids belonging to the Seinajoki family.

\section{Conclusions}

In this work we have shown that Ceres strongly affects the orbital motion of at least 
three asteroid families. So far we know three possible mechanism, involving Ceres, 
that may be at work here. These are close encounters, the 1/1 mean motion 
resonance and the linear secular resonance with (1) Ceres.
However, as we found that in all three cases most of the evolution
is taking place within a narrow range of the semi-major axis, the first two mechanisms
seem to be very unlikely. This range does not correspond to the location of the
1/1 resonance with Ceres, and there is no reason that close encounters affect only
objects within this specific range of semi-major axes. Thus, the only plausible 
explanation is that linear nodal secular resonance with Ceres
is causing the dispersion of orbital inclinations. 

The results presented here have demonstrated that the linear secular resonance with the
dwarf planet Ceres is the mechanism that may perturb orbits of some asteroids to
a very high degree. In particular, the $\nu_{1c}=s-s_c$ resonance is the
perturber of the Hoffmeister, Astrid and Seinajoki asteroid families.
Therefore, this new player in the field definitely should be considered in any future
study of the dynamical evolution of asteroid families.

\section*{Acknowledgments}

We would like to thank the referee, prof. Valerio Carruba, for his valuable comments which 
helped to improve the manuscript. 
The work of B.N., G.T., and S.M. has been supported by the European Union [FP7/2007-2013], 
project: "STARDUST-The Asteroid and Space Debris Network". 
Numerical simulations were run on the PARADOX-III cluster
hosted by the Scientific Computing Laboratory of the Institute of Physics Belgrade.

\end{document}